\documentstyle[epsfig]{mn}

\newif\ifAMStwofonts

\def\etal{et al.}

\ifoldfss
  \ifCUPmtlplainloaded \else
    \NewTextAlphabet{textbfit} {cmbxti10} {}
    \NewTextAlphabet{textbfss} {cmssbx10} {}
    \NewMathAlphabet{mathbfit} {cmbxti10} {} 
    \NewMathAlphabet{mathbfss} {cmssbx10} {} 
  \fi
  \ifAMStwofonts
    \ifCUPmtlplainloaded \else
      \NewSymbolFont{upmath} {eurm10}
      \NewSymbolFont{AMSa} {msam10}
      \NewMathSymbol{\upi}     {0}{upmath}{19}
      \NewMathSymbol{\umu}     {0}{upmath}{16}
      \NewMathSymbol{\upartial}{0}{upmath}{40}
      \NewMathSymbol{\leqslant}{3}{AMSa}{36}
      \NewMathSymbol{\geqslant}{3}{AMSa}{3E}

      \let\leq=\leqslant 
       
    \fi
  \fi
\fi 

\ifnfssone
  \newmathalphabet{\mathit}
  \addtoversion{normal}{\mathit}{cmr}{m}{it}
  \addtoversion{bold}{\mathit}{cmr}{bx}{it}
  \newmathalphabet{\mathbfit} 
  \addtoversion{normal}{\mathbfit}{cmr}{bx}{it}
  \addtoversion{bold}{\mathbfit}{cmr}{bx}{it}
  \newmathalphabet{\mathbfss} 
  \addtoversion{normal}{\mathbfss}{cmss}{bx}{n}
  \addtoversion{bold}{\mathbfss}{cmss}{bx}{n}
  \ifAMStwofonts
    \ifCUPmtlplainloaded \else
      \UseAMStwoboldmath
      \makeatletter
      \new@mathgroup\upmath@group
      \define@mathgroup\mv@normal\upmath@group{eur}{m}{n}
      \define@mathgroup\mv@bold\upmath@group{eur}{b}{n}
      \edef\UPM{\hexnumber\upmath@group}
      \new@mathgroup\amsa@group
      \define@mathgroup\mv@normal\amsa@group{msa}{m}{n}
      \define@mathgroup\mv@bold\amsa@group{msa}{m}{n}
      \edef\AMSa{\hexnumber\amsa@group}
      \makeatother
      \mathchardef\upi="0\UPM19
      \mathchardef\umu="0\UPM16
      \mathchardef\upartial="0\UPM40
      \mathchardef\leqslant="3\AMSa36
      \mathchardef\geqslant="3\AMSa3E

      \let\leq=\leqslant 

    \fi
  \fi
\fi 

\ifnfsstwo
  \DeclareMathAlphabet{\mathbfit}{OT1}{cmr}{bx}{it}
  \SetMathAlphabet\mathbfit{bold}{OT1}{cmr}{bx}{it}
  \DeclareMathAlphabet{\mathbfss}{OT1}{cmss}{bx}{n}
  \SetMathAlphabet\mathbfss{bold}{OT1}{cmss}{bx}{n}
  \ifAMStwofonts
    \ifCUPmtlplainloaded \else
      \DeclareSymbolFont{UPM}{U}{eur}{m}{n}
      \SetSymbolFont{UPM}{bold}{U}{eur}{b}{n}
      \DeclareSymbolFont{AMSa}{U}{msa}{m}{n}
      \DeclareMathSymbol{\upi}{0}{UPM}{"19}
      \DeclareMathSymbol{\umu}{0}{UPM}{"16}
      \DeclareMathSymbol{\upartial}{0}{UPM}{"40}
      \DeclareMathSymbol{\leqslant}{3}{AMSa}{"36}
      \DeclareMathSymbol{\geqslant}{3}{AMSa}{"3E}

      \let\leq=\leqslant 

    \fi
  \fi
\fi 

\ifCUPmtlplainloaded \else
  \ifAMStwofonts \else 
    \def\upi{\pi}
    \def\umu{\mu}
    \def\upartial{\partial}
  \fi
\fi

\title{Annihilation Radiation from a Dark Matter Spike at the Galactic Centre}
\author[G. Bertone, G. Sigl and J. Silk]{Gianfranco Bertone$^{1,2}$, G\"unter Sigl$^1$, Joseph Silk$^2$ \\
1. Institut d'Astrophysique, F-75014 Paris, France \\       
2. Department of Astrophysics, University of Oxford, NAPL 
\\ Keble Road, Oxford OX13RH, United Kingdom}
\date{}

\pagerange{\pageref{firstpage}--\pageref{lastpage}}
\pubyear{1994}

\begin{document}

\maketitle

\label{firstpage}

\begin{abstract}
We study annihilation radiation of neutralinos
in the Galactic centre, assuming the existence of a 'spike' 
in the dark matter density 
profile, due to adiabatic accretion onto the massive black hole 
lying at the Galactic centre. Under this assumption we find that 
it is possible to reproduce 
the observed SgrA* emission at radio and gamma-ray frequencies in 
a consistent scenario with a magnetic field close to the
equipartition strength
and with values of $\gamma$ (the density profile power law index) 
around 0.1. 

\end{abstract}


\section{Introduction}
In a recent paper, Gondolo and Silk (2000) showed that the existence 
of a massive black hole in the Galactic centre would produce a 'spike' 
in the dark matter density profile due to the adiabatic accretion of 
dark matter particles.  It has
 subsequently been argued that such a spike may be destroyed by mergers
(Ullio, Zhao  and Kamionkowski, 2002;
Merritt et al., 2002). We point out here that such
dynamical destruction processes are unlikely to have occurred for the Milky Way.
However our knowledge of black hole formation is so uncertain that we cannot
make this assertion with any real confidence,
and we therefore propose a new means of observing the central spike via gamma ray
emission. Both the  flux and spectrum of the galactic centre annihilation gamma ray
source are consistent with EGRET observations (as long as the initial cusp profile is
rather shallow, $\propto r^{-\gamma}$ with $\gamma\approx 0.1$) 
and if this interpretation is correct,
will potentially  enable us to deduce the magnetic field in the vicinity of SgrA*. 

Supermassive black holes (SMBH) are ubiquitous in galactic nuclei.
Their formation remains a mystery. However circumstantial evidence points strongly
towards formation in the highly dissipative environment of the forming galaxy.
This is based on the remarkably tight correlation between black hole mass and bulge
velocity dispersion
(Ferrarese and Merritt, 2000;
Gebhardt et al., 2000; Merritt and Ferrarese, 2002) that
attests to a common origin, which, given the
fact that our bulge formation preceded that of the disk some 12 Gyr ago,
places  black hole formation at an epoch when our galaxy was extremely gas-rich.
Now dynamical formation of the SMBH by stellar  collisions and mergers
is an extremely slow process, as also would be formation by black hole mergers.
Theory supports a dissipative origin, and the protogalactic origin
provides the ideal environment.  Thus, the SMBH formed early in the protogalaxy,
which in turn condensed in a dark halo
of weakly interacting particles that constitute  cold dark matter (CDM).
 
A spike would inevitably have developed in the CDM halo core as the SMBH formed by
dissipation of gas infall. The critical issue is: were there subsequent mergers that
destroyed the spike and boosted the SMBH mass?
The answer is yes and no.
Yes for ellipticals and for early type galaxies with massive bulges.
But no for the Milky Way, and presumably for similar late-type galaxies with small
bulge-to-disk ratios.
 
There is little doubt that massive spheroids formed by mergers,
and most likely by a series of mergers.
Observations points to a drastic increase in the merger rate associated with
starbursts in the early universe, and the inferred formation rate of dust-shrouded
spheroids. The phenomenological case for merger-driven spheroid formation is strong
(Sanders and Mirabel, 1996).
Theory, for those who take it seriously, indicates that mergers were common in the past
(e.g. Steinmetz and Navarro, 2002).
SMBH fit into the merger scenario by virtue of the fact that a major merger provides a
dramatic means for channeling gas into the central regions of the merged galaxy (e.g.
Barnes and Hernquist, 1996),
thereby instigating the formation of the SMBH. Possibly later mergers
enable the central SMBH to merge and thereby
develop and maintain the observed correlation between SMBH mass and bulge central
velocity dispersion.
 
But what can be stated with  considerable confidence is that our Milky Way galaxy
underwent one significant merger
12 Gyr ago. This resulted in the formation of the bulge, and presumably therefore of
the SMBH, and of the thick
disk. Gas disk formation is a consequence of a gas-rich merger,
and such disks commonly characterize merger remnants
(Barnes, 2002).
The chemical evidence for a unique merger origin in the case
of our Milky Way's thick disk is compelling (Wyse, 2001).
The thick disk amounts to as much as 20 percent of the thin disk, so that this one
merger was indeed a major merger. But the continuity between thin disk, thick disk, and
bulge would have been destroyed had anything significant happened since in the way of a merger.
In particular, the bulge was formed then, and therefore the SMBH formed shortly afterwards if
we accept a dissipative protogalactic origin.
The spike developed in the halo core when the SMBH formed by gas infall, and
 no subsequent  major dynamical merging event occurred that could have destroyed it.
 
We note that an alternative   theory for bulge formation in Milky
Way-type galaxies  has no recourse at all to a merger (e.g. Emsellem, 2001).
Gravitational instability of a cold disk
and secular evolution
is capable of forming the bulge, and such a process would also be capable of driving in
the gas to form the SMBH. Only small bulges are
presumably formed in this manner, the massive bulges forming via mergers.
Hence theory  and phenomenology of the Milky Way formation
support our contention
 that the CDM spike presents a primordial feature that developed as the SMBH formed 
and has not subsequently been destroyed.
 
Other dynamical effects, such as the influence on the dark
halo of the formation of the galactic stellar nucleus bulge
disc (Klypin, Zhao \& Somerville,2001), have been neglected 
here. The consequent enhancement of dark matter density in 
the Galactic centre would further reduce the portion of 
dark matter parameter space we showed to be consistent 
with observations.

The SMBH acts as an amplifier  of the CDM,
locally enhancing the annihilation rate.
The enhancement of the
particle number density and self-annihilation rate in
the proximity of
the central black hole provides a means of probing the nature of CDM by indirect
detection of
the various annihilation signatures.
There are three such signatures: high energy neutrinos, synchrotron emission
from high energy positrons and electrons, and high energy gamma rays.
We have previously discussed the observability of neutrinos and of synchrotron emission
from the spike  at the GC. Here we discuss the gamma ray signature
for
 our favoured dark matter candidate, the so-called 
neutralino, arising in the framework of supersymmetric 
theories (for an extensive review of dark matter candidates see 
e.g. Bergstrom, 2000).

In particular, we have previously  estimated
(Bertone, Sigl and Silk, 2001) the corresponding enhancement of 
annihilation signals, such as synchrotron radiation originating 
from the propagation of e+e- pairs in the galactic magnetic field.
We were  able to constrain neutralino mass and cross section, the 
constraints being strongly dependent on the slope of the 
density profile. Here we extend this analysis to continuum gamma-ray emission, 
which, in absence of the central spike, has previously been evaluated 
by several authors (see e.g. Berezinsky et al., 1994).
  
When considering neutralinos annihilating in a halo with a power-law 
density profile, even with an index as high as 1.0, a deficit of gamma-rays 
is found, with respect to the flux measured by EGRET (see 
Mayer-Haesselwander, 1998).  An explanation for this excess has been 
proposed in terms of 'clumpiness' of the halo by Bergstrom et al. (1999). However rather extreme clumpiness factors are required, and it is not
clear that the halo substructure has survived
at least as judged by counting the number of dwarf satellites and by theoretical
arguments pertaining to excessive
loss  of angular momentum by clumpy protogalactic gas
(e.g. Silk, 2001).
 
In the present case, the gamma-ray flux is enhanced by the presence 
of the central spike at the Galactic Centre.
We will  show that a consistent scenario can 
be developed  that simultaneously reproduces 
the observed SgrA* radio spectrum as well as the high energy gamma ray emission
observed from the  Galactic Centre by EGRET
if the initial 
cusp profile, before adiabatic accretion onto the SMBH, is 
rather shallow, with power law index $\gamma \approx 0.1$.

Although some astrophysical processes 
may flatten considerably 
the cusps respect to conventional $\gamma \approx 1$ profiles
(see Katz \& Weinberg, 2002 and Valenzuela \& Kyplin, 2001) and lead
to the formation of shallow cusps, we prefer here to treat $\gamma$
as an unknown parameter.

Alternative models exist in literature, to explain the spectrum
of SGR A*, suggesting that high energy emission could have an origin
in pulsars or around a supermassive black hole (e.g.  
Mayer-Haesselwander et al., 1998).
 
In section 2 we give details about the  supersymmetric scenarios 
used to make detailed estimates of expected fluxes. 
Section 3 is devoted to the description of the halo model with
a central spike. The detailed analysis of synchrotron and gamma-ray 
emission is presented in section 4. The final section contains
results and conclusions.
 
\section[]{Neutralinos as dark matter candidates}
There is convincing evidence that a large part of
the dark matter 
is non-baryonic. Here we focus on the lightest supersymmetric 
particle, which in most supersymmetric scenarios turns out 
to be the neutralino  (e.g. Jungman et al., 1996, and references therein).

One of the best reasons to consider neutralinos as dark matter 
candidates is that SUSY was not introduced to solve cosmological or
astrophysical problems:  rather,
its motivations are related to
some fundamental problems of theoretical physics, such as the 
fundamental difference between bosons and fermions, the mass
hierarchy problem and Grand Unification (e.g. Jungman et al., 1996).
  
Despite its success in solving some of these problems, it is 
still not very  predictive, containing a huge number of free 
parameters. This is mainly due to the fact that SUSY is not 
an exact symmetry of nature, and should be broken at some scale
(otherwise we would observe superpartners of ordinary particles 
with the same masses, and we do not).
  
It is possible, however, to reduce the number of free parameters 
by  making several reasonable assumptions. The definition of a SUSY
scenario thus  corresponds to specifying a set of assumptions
that aim
to reduce and constrain the parameter space. The results discussed in this paper are 
obtained in the framework of the Minimal Supersymmetric 
Standard Model (see Haber\&Kane, 1985, or Jungman et al, 1996) as 
implemented by Gondolo \textit{et al} (2000) in the DarkSUSY 
code, in which there are seven free 
parameters: one scalar mass parameter $m_0$, the mass of
the pseudo-scalar 
Higgs $m_A$, one universal gaugino mass $M_2$, $\rm{tan}\beta$ (ratio of 
the VEVs of the two Higgs fields) and  two trilinear parameters 
$A_b$, $A_t$.
 
We run extensive scans 
of the dMSSM 7-dimensional parameter space and choose the models 
consistent with accelerator and cosmological constraints. In 
particular, all the models discussed here have a relic density 
$0.025 \leq \Omega h^2 \leq 0.3$.
 \section{Modified density profile}
 
As discussed by Gondolo and Silk (2000), the presence 
of a black hole can dramatically modify the dark matter density 
profile near the Galactic Centre.
 \begin{figure}
\epsfig{file=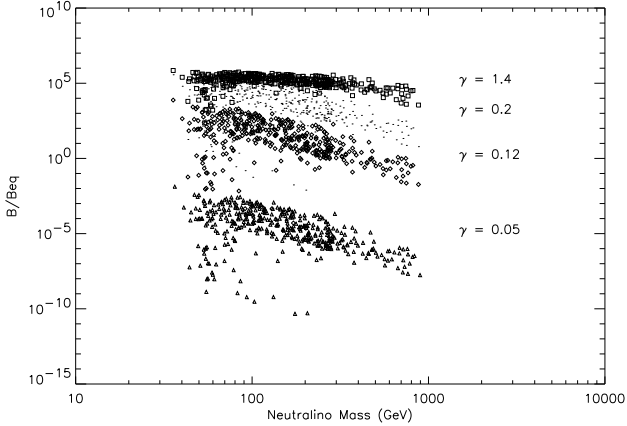,width=84mm}
 \caption{Required values of the magnetic field, relative to
the equipartition field (see eq. \ref{mag}) to reproduce the observed 
spectrum of SgrA* at a frequency of $408\rm{MHz}$. Each point represents a 
different supersymmetric ``model'' consistent with cosmological 
relic abundance and accelerator constraints.}
\label{synchro} 
\end{figure}

Starting from an ordinary power law for the dark matter 
density profile in the Galaxy (as predicted by N-body simulations), 
one obtains, after adiabatic accretion onto the central black hole, 
the following 'modified' profile 
\begin{equation} 
\rho' = \left[ \alpha_{\gamma} \left( \frac{M}{\rho_D 
D^3}\right)^{3-\gamma} \right]^{\gamma_{sp}-\gamma} \rho_D \; g(r) \left( 
\frac{D}{r} \right)^{\gamma_{sp}} 
\label{modaccre} 
\end{equation} 
where $\gamma_{sp}=(9-2\gamma)/(4-\gamma)$, $D$ is the solar distance from 
the Galactic Centre and $\rho_D=0.24 GeV/c^2/cm^3$ is the density in the 
solar neighborhood. 
  
The factors $\alpha_\gamma$ and $g_\gamma(r)$ cannot 
be determined analytically (for approximate expressions and numerical 
values see paper I). The expression~(\ref{modaccre}) is only valid in a 
central region of size $R_{sp}=\alpha_\gamma D (M/\rho_D 
D^3)^{1/(3-\gamma)}$ where the central black hole dominates the 
gravitational potential.
  
Self-annihilation of dark matter particles further modifies the density 
profile: if $t_{BH}$ is the age of the SMBH, the maximal dark matter 
density is 
\begin{equation} 
\rho_{core}=m_x/\sigma v t_{BH} 
\label{roco} 
\end{equation} 
where $m_x$ and $\sigma v $ are respectively the neutralino mass 
and annihilation cross section (times relative velocity).
  
The final profile is given by 
\begin{equation}
\rho_{dm}(r)= \frac{\rho'(r) \rho_{core}}{\rho'(r)+\rho_{core}} 
\end{equation}
following a power law for large values of $r$, and with a flat core of 
density $\rho_{core}$ and  dimension 
\begin{equation}
R_{core}=R_{sp} \left( \frac{\rho(R_{sp})}{ \rho_{core}} \right) 
^{(1/\gamma_{sp})}. 
\end{equation} 
This result is obtained neglecting dynamical effects of hierarchical 
merging. If one takes into 
account the former effects, then 
shallower dark matter profiles are formed, with power law index as 
low as $\gamma \approx 0.5$ or even  flatter (see Merritt et al., 2002, Nakano 
\& Makino, 1999, and Ullio, Zhao \& Kamionkowski, 2001).
 
However, as stressed in the introduction, post-merger black hole formation, as seems to
be the case for our galaxy,
means that the spike will still form, albeit with  a reduced value of
$\gamma_{sp}$ that however is always larger than 7/4 and hence gives a
large annihilation flux enhancement.
An interesting effect arises if the massive black hole does not lie
exactly at the centre of the dark matter halo, as may be the consequence of being
embedded in a central massive star cluster of size
$a< 0.01$pc. According to
 Chatterjee, Hernquist and Loeb (2002),
for solar mass stars,
stochastic motions of the SMBH occur
in a harmonic potential of scale $\sim  10^{-3} a$pc,
and this would  have the effect of further
softening the spike; note that annihilation softening occurs
at  $\sim 10^{-6} $pc (Gondolo and Silk 1999)).

\section{Annihilation radiation}
 
The neutralino self-annihilation rate is very sensitive to the  neutralino
number density, namely
 
\begin{equation} 
\Gamma = \frac{\sigma v}{m_x^2} \int_0^{\infty} \rho_{dm}^2 4 \pi r^2 \;\; 
dr\,, 
\end{equation}

We therefore expect the annihilation signals coming from the neighborhood 
of the central black hole to greatly exceed 
the signals from other regions of the halo.
  
Among secondary products of neutralino annihilation, we study high 
energy photons, originating via neutral pion decays, and the synchrotron 
radiation of electron-positron pairs (originating from the decays
of charged pions) in the Galactic magnetic field.

\subsection{Synchrotron radiation}
 \begin{figure}
\epsfig{file=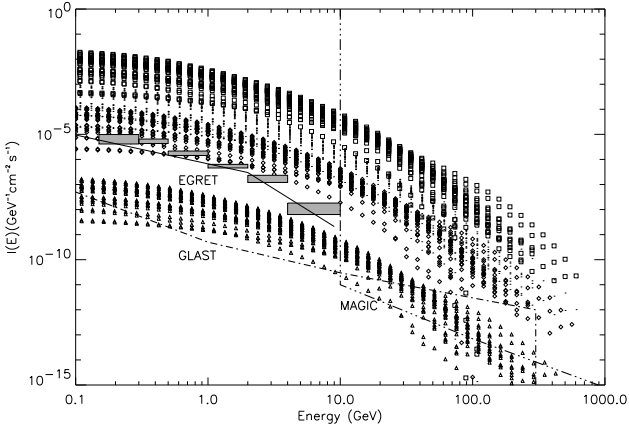,width=84mm}
\caption{Expected gamma-ray flux for the same set of supersymmetric
models and $\gamma$=0.05(triangles), $\gamma$=0.12(diamonds), $\gamma$=0.2 (dots), $\gamma$=1.0 (squares), along with EGRET data
(Narayan et al, 1998) and expected sensitivities for GLAST (1 month observation time) and MAGIC (50 hours).}
\label{gammas}
\end{figure}
 
In the region of interest, the Galactic magnetic field can be 
roughly approximated using the 'equipartition assumption' 
(see Melia, 1992), such that 
\begin{equation} 
B_{eq}(r) = 1\mu G \left( \frac{r}{pc} \right)^{-5/4}. 
\label{mag} 
\end{equation}
This value is obtained assuming equality between magnetic and gravitational 
energy density. In the case of accretion with conservation of mass flux 
$\rho v r^2=$constant, 
hence, for keplerian velocities, $rho \propto r^{-3/2}$ and the gravitational 
energy $\epsilon_{GR}\propto r^{-5/2}$ (for more details see 
also Shvartsman, 1971). 

In paper I, we estimated the synchrotron luminosity at a given
frequency $\nu$ in such a magnetic field to be 
\begin{equation} 
L_{\nu} \approx \frac{9}{8} \left(\frac{1}{0.29 \pi} \frac{ m_e^3 c^5}{e} 
\right)^{1/2} \frac{\Gamma Y_e(>E)}{\sqrt{\nu}} \; I 
\label{synchrolum}
\end{equation} 
where 
\begin{equation} 
I = \int_0^{\infty} dr \;\; 4 \pi r^2 f_e(r) B^{-1/2}(r) 
\label{syn}
\end{equation} 
and the function $f_e(r)$ is defined as 
 \begin{equation} 
f_e(r)= \frac{\rho_{sp}^2}{  \int_0^{\infty} \rho_{sp}^2 4 \pi r^2 \;\; 
dr} 
\end{equation} 

Eq. \ref{synchrolum} may appear to be counterintuitive, due to the presence
of the factor $B^{-1/2}$ in the integral defined in eq. \ref{syn}.
Indeed an increase of the magnetic field would produce a decrease
of the expected flux for a given frequency, simply because the full
spectrum would be 'shifted' towards higher energies. The synchrotron flux 
(see e.g. Bertone, Sigl and Silk, 2001) is a growing function of 
frequency, roughly following a power law, until the cut-off frequency,
which is the critical synchrotron frequency (see Rybicki  \&  Lightman, 1979)
of photons produced from the highest energy electrons (of energy 
$\approx m_x$). The cut-off frequency can be expressed as 
\begin{equation} 
\nu \approx 100 \left(\frac{\rm{B}}{\mu G}\right) \left(\frac{m_x}{100 \rm{GeV}}\right) \rm{GHz}
\end{equation} 

An increase of the magnetic field would thus move  the photons originated 
by the same electron population towards higher frequencies,
thus reducing the flux for any given frequency, but extending it
at higher frequencies, due to the increase of the cut-off frequency. 
 \begin{figure}
\epsfig{file=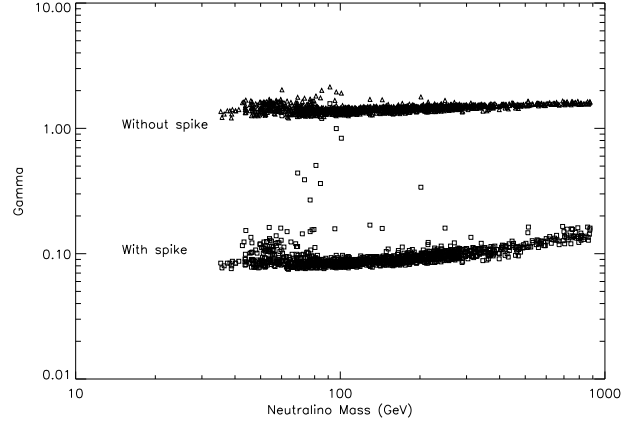,width=84mm}
\caption{Required value of $\gamma$ to reproduce the EGRET data normalization for the same set of supersymmetric models as in figure 1. Squares and triangles are for a halo model respectively with or without the central spike.}
\label{reqgamma}
\end{figure}

To obtain the observed radiation, one should multiply the luminosity 
$L_\nu$ for the synchrotron self-absorption coefficient, defined as 
(see Rybicki  \&  Lightman, 1979) 
\begin{equation} 
A_{\nu}= \frac{1}{a_{\nu}} \int_{0}^\infty (1-e^{-\tau(b)}) 2 \pi b \;\; 
db 
\label{anu} 
\end{equation}
where $\tau(b)$ is the optical depth as a function of the cylindrical 
coordinate b
 
\begin{equation}
\tau(b)= a_{\nu} \int_{d(b)}^{\infty} f_e(b,z) \;\;dz
\label{la}
\end{equation}
and the coefficient $a_{\nu}$ in our case is given by
\begin{equation}
a_{\nu} =\frac{\Gamma Y}{4 \pi} \frac{c^2}{\nu^3},
\end{equation}
(see paper I for details about lower limit of integration in
eq.\ref{la} and approximations introduced).

We found that synchrotron self-absorption can reduce the predicted radio flux
by several
orders of magnitude .
We evaluated the synchrotron luminosity for a wide set of
supersymmetric models (using the DarkSUSY code)
and for different values of $\gamma$. We used, in particular,
the expected luminosity at $\nu=408MHz$ to estimate the
magnetic field required to reproduce the observed flux.

In figure \ref{synchro} we show the value of the magnetic field,
relative to the equipartition field of eq. \ref{mag} for
 a wide set of supersymmetric model and 4 different values of
 $\gamma$.

A value of $\gamma$ between 0.05 and 0.1 seems to reproduce
the right normalization at radio wavelengths,
if combined with a magnetic field close to equipartition. Other values
of the density profile power law index would instead require
unacceptable values of B.

\subsection{Gamma-ray emission}
As discussed above, neutralino annihilation produce a continuum
$\gamma$-ray spectrum, originating from the decay of neutral pions,
which in turn come from the hadronization process of quark-antiquark
pairs.

Referring to the paper of Gondolo and Silk (1999) we write the
contribution of the spike as
\begin{equation}
\label{phispike2}
\Phi^{\rm spike} =
\frac{ \rho_D^2 Y_{\gamma} \sigma v D} {m_x^2} \, 
\left( \frac{ R_{\rm sp} }{ D } \right)^{3-2\gamma} \, 
\left( \frac{ R_{\rm sp} }{ R_{\rm in} } \right)^{2\gamma_{\rm sp}-3} , 
\end{equation} 
where $R_{\rm in} = 1.5 \left[ (20 R_{\rm S})^2+R_{\rm core}^2 
\right]^{1/2} $ and $Y_{\gamma}$ is the number of photons produced 
per annihilation.

This formula is valid in the case of adiabatic accretion 
from an initial power law density profile. 
If one considers an initial {\it isothermal} profile (i.e. a profile 
with a flat central  core), the enhancement of the flux in the 
Galactic Center would be negligible compared to radiation from 
annihilations along the line of sight (see Gondolo and Silk, 1999).
  
We show in figure \ref{gammas} the expected fluxes for a 
subset of supersymmetric models and the same values of $\gamma$ as 
in figure \ref{synchro}. A value of $\gamma$ between 0.05 and 
0.1 can reproduce the normalization of the observed flux. This is 
fully consistent with the result found in the previous section.

Larger values of $\gamma$ are also able to reproduce the 
EGRET normalization, as shown in figure \ref{reqgamma}. In
particular we found that even if most of the models would
require a $\gamma$ of the order of 0.1 or lower, some of 
them could go as high as $\gamma \approx 1$. Those models
correspond to very low predicted fluxes, and   
are thought to require an important level of fine-tuning 
of the seven input parameters. For comparison
we also show the expected results in a scenario without the
central spike, in which case a range of $\gamma$ between 
1 and 2 is required. 

\begin{figure}
\epsfig{file=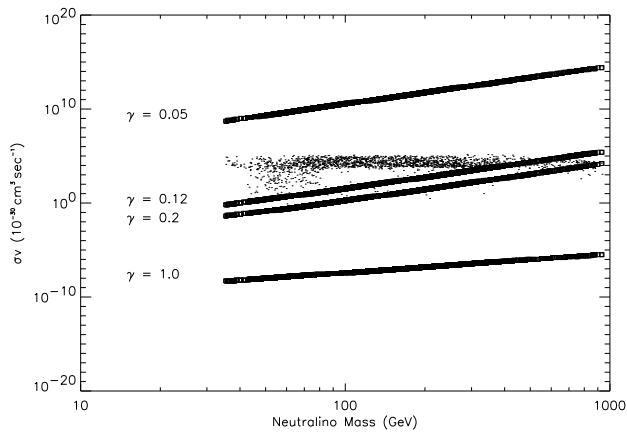,width=84mm}
\caption{Squares: required annihilation cross section to reproduce the EGRET normalization for different values of $\gamma$ assuming profiles with a central spike. We show for comparison (dots) the total annihilation cross section for all SUSY models.}
\label{cross}
\end{figure}

Finally in figure \ref{cross} we show the required cross sections
to reproduce the EGRET data normalization for all the supersymmetric
models and four different values of $\gamma$ (in the case of presence 
of the central spike). Here again it is possible to read out the range 
of values of $\gamma$ which can reproduce the observed EGRET normalization 
with cross sections close to those predicted in our supersymmetric scenario.  
 
\section{Results and Conclusions}
   
In the framework of a halo model with a spike around the 
central black hole, we showed that a 
consistent scenario can be built, reproducing at the same 
time both the radio and gamma-ray emission.
  
Observed radio emission can be explained by  synchrotron 
emission of secondary e-e+pairs in the Galactic magnetic 
field. The enhancement of annihilation rate due to the central 
spike and synchrotron self-absorption 
were the two main ingredients of our calculation. 
  
Despite  the lack of distinctive features of continuum gamma-ray 
emission, we find  full consistency of the results obtained 
with  the radio emission, the EGRET flux normalization being 
reproduced with  values of $\gamma \approx0.1$, when assuming
 profiles with a central spike.
  
The argument can also be turned the other way round and 
interpreted as a 'measure' of the 
Galactic magnetic field: we can in fact decide to select the 
values of $\gamma$ reproducing the normalization of the 
observed gamma-ray emission (figure \ref{reqgamma}), and go 
back to figure \ref{synchro} to read out the corresponding 
value of $B^*$, which for most of the models is indeed of 
order the equipartition value, thus confirming the consistency 
of our scenario.

Forthcoming experiments, such as GLAST, will probe energies well 
above the EGRET measurements, up to 300 GeV. A sharp cutoff 
around the neutralino mass is predicted for this scenario, 
if the annihilation radiation gives the dominant contribution 
to normalization. Furthermore other 'smoking gun' signatures, 
such as narrow high energy gamma ray lines  could give 
further information and constraints on neutralino properties 
or halo profiles.

\section*{Acknowledgments}

We   would like to thank J.Edsjo for enlightening discussions
and help on the use of the DarkSUSY code. This activity was carried
out in the framework of the 'Supersymmetry and Early Universe Network'
(HPRN-CT-2000-00152).

\bsp

\label{lastpage}


\begin{thebibliography}{}

 \bibitem{barn} Barnes, 2002, astro-ph/0201250

\bibitem{bar} Barnes and Hernquist, 1996, ApJ 471, 115 

\bibitem{ber}  Berezinsky V., Bottino A., Mignola G., 1994 Phys.Lett. 
B325, 136


\bibitem{bergstrom} Bergstrom L., 2000, Rept.Prog.Phys. 63, 793

\bibitem{bergstrom2} Bergstrom L., 1999, Edsjo J. and Gondolo P.,
Phys.Rev. D59, 043506

\bibitem{bertone} Bertone G., Sigl G. and Silk J., 2001, Mon.Not.Roy.Astron.Soc., 
326, 799

\bibitem{chat} Chatterjee, Hernquist and Loeb, 2002, ApJ in press, 
astro-ph/0107287 

\bibitem{emsellem} Emsellem 2001, astro-ph/0109142

\bibitem{ferra} Ferrarese L. and Merritt D., 2000, ApJ 539, L9

\bibitem{geb} Gebhardt \etal, 2000, ApJ, 539, L13


\bibitem{gondolo} Gondolo P. \& Silk J., 1999, PRL 83, 1719
 
\bibitem{Gondolo:2000ee}
 
P.~Gondolo, J.~Edsjo, L.~Bergstrom, P.~Ullio and E.~A.~Baltz, 
astro-ph/0012234. 

\bibitem{Haber:1984rc}
H.~E.~Haber and G.~L.~Kane, 1985
Phys.\ Rept.\  {\bf 117}, 75.

\bibitem{Jungman:1995df} Jungman G., Kamionkowski M. and Griest K., 1996
Phys.\ Rept.\  {\bf 267}, 195

\bibitem{klypi} Klypin A., Zhao H.S., Somerville R.S., astro-ph/0110390 

\bibitem{klypi2}

\bibitem{mayer} Mayer-Haesselwander H.A. \etal, 1998, A\&A 335, 161

\bibitem{melia} Melia F., 1992,  ApJ 387, L25

\bibitem{merritt} Merritt D. \etal , astro-ph/0201376

\bibitem{merritt2} Merritt D. and Ferrarese L., 2002, astro-ph/0107134

\bibitem{Nakano:1999ta}
 
Nakano T. and ~Makino J., ApJ 510, 155
 
 
 
\bibitem{spectrum}  Narayan et al., 1998, ApJ 492, 554
 
\bibitem{Rybicki} Rybicki G.B., Lightman A.P., Radiative Processes in 
Astrophysics, 1979, John Wiley \& Sons
 
\bibitem{san} Sanders and Mirabel, 1996, ARA\&A 34, 749

\bibitem{shva} Shvartsman V.F., 1971, Soviet Astronomy-AJ, 15, 377

\bibitem{silk} Silk J., 2001, astro-ph/0112295


\bibitem{stein} Steinmetz and Navarro 2002, astro-ph/0202466

\bibitem{Ullio:2001fb} Ullio P., Zhao H.and Kamionkowski M., 
astro-ph/0101481.

\bibitem{vale} Valenzuela O. \& Klypin A., astro-ph/0204028 

\bibitem{katz} Weinberg M.D. \& Katz N., astro-ph/0110632 

\bibitem{wyse} Wyse R.F.G., 2000, astro-ph/0012270
 

\end{thebibliography}
\end{document}